\documentclass[review,12pt]{elsarticle}

\usepackage{graphicx}
\usepackage{color}
\usepackage{url}
\usepackage{amsmath}
\usepackage{amssymb}
\usepackage{dcolumn}
\usepackage{afterpage}

\newcommand{\s}{\sum}

\newcommand{\beq}{\begin{equation}}
\newcommand{\eeq}{\end{equation}}

\newcommand{\Rg}{R_g}
\newcommand{\kf}{k_f}
\newcommand{\df}{d_f}
\newcommand{\Rout}{R_{\textrm{out}}}
\newcommand{\Rgeo}{R_{\textrm{geo}}}

\biboptions{comma,square}

\journal{Journal of Colloid and Interface Science}

\begin{document}

\begin{frontmatter}

\title{Morphology and mobility of synthetic colloidal aggregates}

\author[jrc,apth]{Anastasios D. Melas\corref{cor}}
\ead{admelas@auth.gr}

\author[lux]{Lorenzo Isella}

\author[apth,cperi]{Athanasios G. Konstandopoulos}

\author[jrc]{Yannis Drossinos}

\cortext[cor]{Corresponding author.}

\address[jrc]{European Commission, Joint Research Centre, I-21027
Ispra (VA), Italy}
\address[apth]{Department of Chemical Engineering, Aristotle University,
GR-54124 Thessaloniki, Greece}
\address[lux]{European Commission, DG Energy, L-2530 Luxemburg, Luxemburg}
\address[cperi]{Aerosol \& Particle Technology Laboratory, CERTH/CPERI, P.O. Box
60361, GR-57001 Thessaloniki, Greece}

\date{\today}

\begin{abstract}

The relationship between geometric and dynamic properties of fractal-like
aggregates is studied in the continuum mass and momentum-transfer regimes.
The synthetic aggregates were generated by a cluster-cluster
aggregation algorithm.
The analysis of their morphological features suggests
that the fractal dimension is a descriptor of a cluster's large-scale
structure, whereas the fractal prefactor is a local-structure indicator. For a
constant fractal dimension, the prefactor becomes also an indicator of
a cluster's shape anisotropy.
The hydrodynamic radius of orientationally averaged aggregates was calculated
via molecule-aggregate collision rates determined from the solution of a Laplace equation.
An empirical expression that relates the aggregate hydrodynamic
radius to its radius of gyration and the number of primary particles is proposed.
The suggested expression depends only on geometrical quantities, being
independent of statistical (ensemble-averaged) properties like the fractal dimension and prefactor.
Hydrodynamic radius predictions for a variety of fractal-like aggregates are in very good agreement
with predictions of other methods and literature
values. Aggregate dynamic shape factors and DLCA individual monomer hydrodynamic
shielding factors are also calculated.

\end{abstract}


\end{frontmatter}

\section{Introduction}

Aerosol and colloidal particles may form complex structures via agglomeration~\cite{ThanasisK}
and flocculation. The morphology and hydrodynamic properties of these structures have been
studied extensively in the literature, e.g., Refs.~\cite{friedlander-book,botet-book}, due to
their numerous technological applications: for example, the mobility of power-law
aggregates influences their size distribution, their precipitation behaviour,
and their agglomeration. Even though many studies have investigated the relationship between
geometric and dynamic properties, the prediction of the
hydrodynamic radius from aggregate structural properties remains elusive.

Forrest and Witten~\cite{Forrest}, in their analysis of the
agglomeration of ultrafine smoke particles,
first suggested that the resulting agglomerates
are power-law objects obeying the scaling law (over a finite size range)
\beq
N=k_{f} \Big( \frac{R_{g}}{R_{1}} \Big)^{d_{f}},
\label{eq:scaling}
\eeq
where $N$ is the number of primary particles that form the aggregate,
$d_{f}$ the fractal (or Hausdorff) dimension, $k_{f}$ the fractal
prefactor (also referred to as lacunarity~\cite{Lapuerta10}
or structure factor~\cite{Gmachowski96}),
$R_{g}$ the radius of gyration, and
$R_{1}$ the radius of the primary particles.
We refer to aggregates satisfying the scaling law Eq.~(\ref{eq:scaling}) as
``power-law" aggregates~\cite{WuFriedlander93} (equivalently, fractal-like or quasi-fractal)
because the scaling law
relation is independent of whether the aggregate has a real scale-invariant
(self-similar) morphology.
The fractal dimension provides a quantitative measure of the
degree to which a structure fills physical space beyond
its topological dimension.
The  fractal prefactor, a parameter whose importance is
increasingly being appreciated~\cite{Gmachowski96,Wu,Koylu95a,Sorensen97},
is an essential ingredient
for a complete description of a power-law aggregate, as suggested by the
scaling law. According to Wu and Friedlander~\cite{Wu} it
is a descriptor of packing of the primary
particles, becoming an indicator of the aggregate local structure.
The radius of gyration is a geometric measure of the spatial mass distribution
about the aggregate center of mass.

The calculation of the Stokes friction coefficient of a fractal-like aggregate,
and consequently of its hydrodynamic radius,
is analytically and computationally demanding as it requires
the solution, analytical or numerical, of the creeping-flow Stokes equations.
The hydrodynamic radius of an aggregate is defined as the radius of a sphere
with the same mobility (or equivalently, the same diffusion coefficient) under identical flow
conditions, ensemble-averaged over many aggregates and orientationally averaged~\cite{Wu}.
Several methods have been proposed to calculate it.

Kirkwood and Riseman~\cite{KR} in their pioneering analysis of the translational
diffusion coefficient of flexible macromolecules derived a purely
geometrical expression for the polymer friction coefficient. The derived
expression depends only on monomer-monomer distances in the chain. Their analysis
was based on a double average of the Oseen tensor, a tensor that
describes the perturbed fluid velocity on a surface due to a point source:
an initial average over the internal  configurations of the chain
is followed by an orientational average.
Hubbard and Douglas~\cite{Hubbard} modified their analysis by
avoiding the configurational pre-averaging approximation, retaining
the angular average of the Oseen tensor. The remaining angular average
corresponds to the physical
average over the orientational Brownian motion of the aggregate. They realized that the
orientationally averaged (spherically symmetric) Oseen tensor
is the free-space Green's function of the Laplace operator. Thus, they concluded that
the orientationally averaged hydrodynamic friction of an arbitrarily shaped
Brownian particle may be obtained from the solution of
a Laplace equation.
Hogan and co-workers in a series of papers~\cite{Gopala,ZhangHogan,Thajudeen} calculated
the so-called Smoluchowski radius, the point mass-transfer analogue of the
hydrodynamic radius, via stochastic simulations of point mass-aggregate collision rate.
Their calculations are, in a sense, equivalent to the discrete stochastic simulations
of the Hubbard and Douglas~\cite{Hubbard} continuum approach.
Filippov~\cite{filippov} avoided the previously described
approximations, at the expense of significant numerical effort,
by developing a full
multipole expansion of the Stokes velocity field to obtain the
fluid stress tensor on the aggregate surface. The friction coefficient
was subsequently calculated by integrating the stress tensor over the
aggregate surface.

In this study we use the methodology introduced and validated by Isella and
Drossinos~\cite{LorenzoJCIS11} who calculated, approximately but accurately,
the friction coefficient and the hydrodynamic radius of straight chains
by solving a Laplace equation with appropriate boundary conditions.
Their approach is similar to the continuum approach of Hubbard and Douglas~\cite{Hubbard} and
the single-particle discrete simulations performed by Hogan and collaborators~\cite{ZhangHogan}.
Its advantages are the numerical solution of a simpler equation and
easy computational implementation.
The method as originally proposed is limited to colloidal aggregates or aerosol
particles where mass and momentum transfer occurs in the so-called continuum
regime. In the continuum transfer regime rarefaction effects,
quantifiable by the Knudsen number, $\textrm{Kn} = \lambda/R_1$ where
$\lambda$ is the gas mean free path, are negligible as
$R_1 \gg \lambda$ ($\textrm{Kn} =0$).

The power-law aggregates we use in this work are synthetic in that
they were generated
by an algorithm that does not simulate a physical agglomeration mechanism.
Instead, the algorithm allows the construction of power-law aggregates with specific properties.
In the following, we study the morphology of these synthetic aggregates in an
attempt to identify the geometrical factors that determine their small- and
large-scale structure. We propose an empirical fit that
relates their dynamical properties (hydrodynamic radius)
to structural properties (radius of gyration).

\section{Hydrodynamic radius of synthetic fractal-like aggregates}
\label{sec:FrictionCoefficient}

\subsection{Methodology}

In the continuum regime the Stokes friction coefficient of a $N$-monomer aggregate is~\cite{friedlander-book}
\beq
f_N = \frac{1}{B_{N}} = \frac{k_B T}{D_N} \equiv 6\pi \mu R_{h},
\label{eq:Stokes}
\eeq
where $B_{N}$ is the aggregate mechanical mobility, $D_N$ the Stokes-Einstein diffusion
coefficient, $k_B$ the Boltzmann constant, $\mu$ the fluid viscosity, and $R_{h}$ the
hydrodynamic radius. Equation~(\ref{eq:Stokes}) defines the aggregate
hydrodynamic radius, which equals the
mobility radius in the continuum regime.

Isella and Drossinos~\cite{LorenzoJCIS11} argued that the ratio of two
aggregate-to-monomer friction coefficients,
and correspondingly of their hydrodynamic radii, is related to
a ratio of two molecular collision rates: the molecular collision rate with the
$N$-aggregate ($K_N$) and the molecular collision rate with a monomer ($K_1$).
Accordingly,
\beq
\frac{f_N}{N f_1} = \frac{K_N}{N K_1} = \frac{R_h}{N R_1}.
\label{eq:CollisionRates}
\eeq
The collision rates may be calculated from the
steady-state molecular diffusion equation
[$\nabla^2 \rho(\textbf{r}) = 0$],
via integrating the molecular diffusive flux $\textbf{J}_N = -D_{g} \, {\mathbf{\nabla}} \rho$
over the aggregate surface, where $D_{g}$ is the molecular diffusion coefficient and $\rho$ the gas density.
The appropriate boundary conditions are total absorption
on the aggregate surface
[$\rho (\textbf{r}_{\rm sur}) = 0$, i.e., neglect of multiple scattering events]
and constant fluid density far away from the aggregate
($\rho \to \rho_{\infty}$ for $|\textbf{r}| \to \infty$).
For a monomer, the molecular collision
rate evaluates to $K_1 = 4 \pi D_g R_1 \rho_{\infty}$.
Thus, the friction coefficient may be determined from the
numerical solution of a diffusion equation.
For the diffusion calculations we used the finite-element
software COMSOL Multiphysics~\cite{comsol}.

\begin{figure}[htb]
\begin{center}
\includegraphics[width=0.675\columnwidth,height=0.50\columnwidth,]{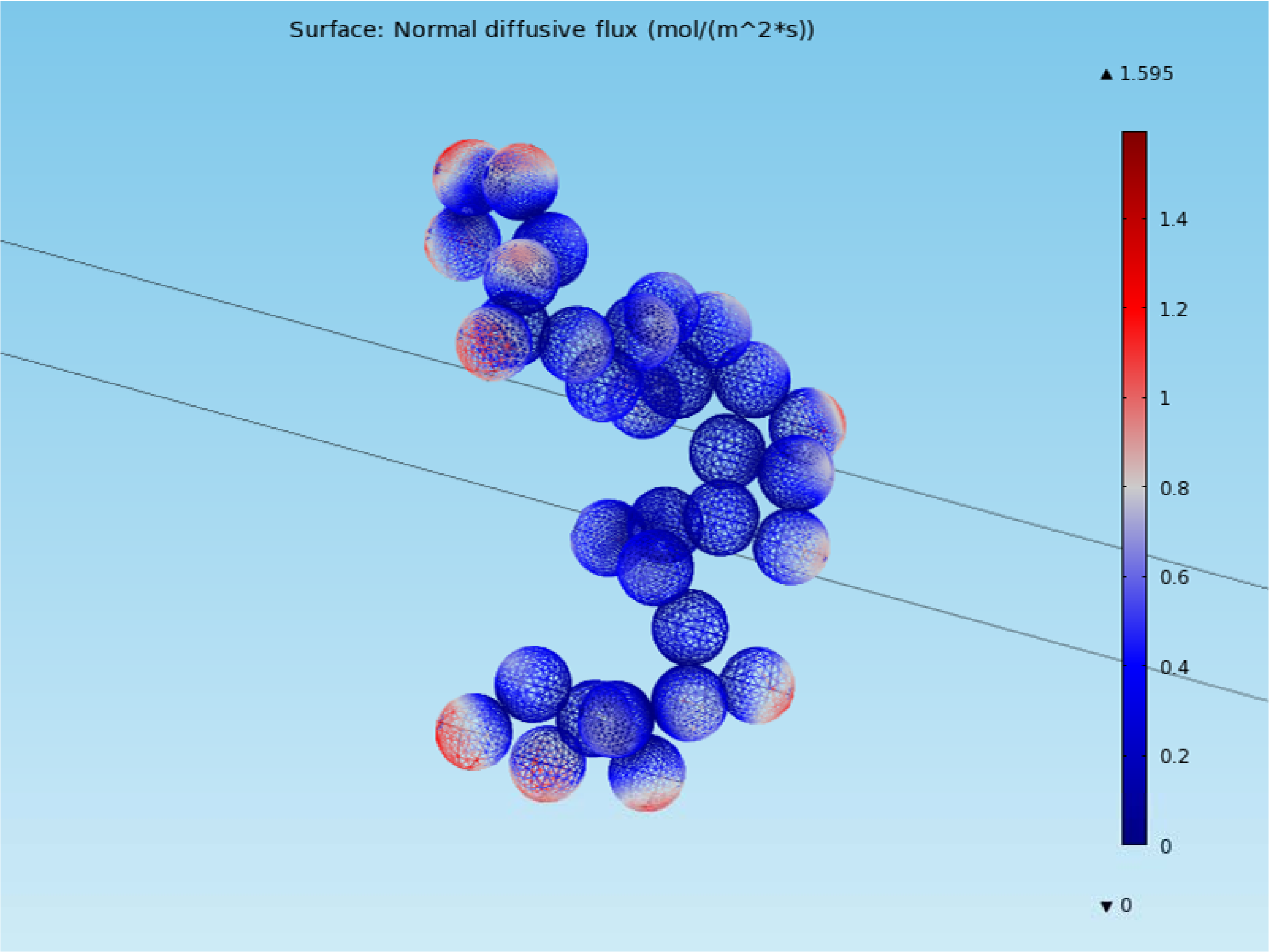}
\caption{Normal diffusive flux on the surface of a DLCA ($d_f=1.8,k_f=1.3)$ synthetic $32$-monomer
power-law aggregate.}
\label{fig:NormalFlux}
\end{center}
\end{figure}

Isella and Drossinos~\cite{LorenzoJCIS11} validated
the methodology for straight chains ($d_{f}=1, k_{f}=\sqrt{3}$) by
solving the diffusion equation in cylindrical coordinates.
We reproduced their calculations in three-dimensional spherical coordinates.
The size of the spherical computational
domain was chosen to be at least two orders of magnitude
larger than a characteristic dimension of the aggregate
to ensure that the condition $\rho_{\infty}=\textrm{constant}$
hold at the computational-domain boundaries.
We also tested the mesh-independence of the solutions.
Figure~\ref{fig:NormalFlux} shows a power-law aggregate with the corresponding normal
diffusive flux, whose integral over the aggregate surface
gives the molecule-aggregate collision rate. The aggregate
hydrodynamic radius is obtained though Eq.~(\ref{eq:CollisionRates}) and the
appropriate normalization via $K_1$.

Furthermore, we validated our calculations for two three-dimensional objects
by comparing them to literature values. We calculated the perpendicular
friction coefficient of two $3$d, symmetric shapes composed of
$8$ particles: a cube and a rectangle. Our results
are compared to the  numerically evaluated, analytical calculations of Filippov~\cite{filippov}
in Table~\ref{table:cube}. The highly accurate collision-rate results provide additional
support that the method is general enough to be extended to power-law
aggregates (with $d_{f}\neq 1$).
\begin{table}[htb]
\caption{Perpendicular friction coefficient of two $8$-monomer symmetric objects.}
\label{table:cube}
\begin{center}
\begin{tabular}{ccc} \hline \hline
Structure & Filippov~\cite{filippov} & Collision rate \\ \hline
Cube ($2\times 2 \times 2$) & 0.293 & 0.290  \\
Rectangle ($2 \times 4 \times 1$) & 0.361 & 0.366\\ \hline \hline
\end{tabular}
\end{center}
\end{table}

\subsection{Generation}

The power-law aggregates used in our simulations were
created with the tunable cluster-cluster aggregation algorithm
proposed by Thouy and Jullien~\cite{Thouy} and modified by
Filippov et al.~\cite{filippov-review}.
The use of a ``mimicking" algorithm, i.e., an algorithm that is  not based on a physical
agglomeration mechanism, allows us to generate aggregates that have
prescribed number of primary particles, fractal dimension, and fractal
prefactor. The synthetic aggregates satisfy exactly the scaling law
by construction. They share many features with
aggregates generated by physical process-based algorithms,
and, more importantly, they provide an ensemble of well characterized aggregates to
investigate the relationship between their static and dynamic properties.

We consider equal-sized, spherical, and non-overlapping monomers
(primary particles).
The creation of a fractal-like object starts by specifying the desired total number
of primary particles $N=2^{n}$ where $n$ is the number of
generations. Initially we create $N/2$ dimers; the dimers
stick together to form $4$-mers by choosing randomly a sticking point and a
sticking angle, a process that guarantees that each aggregate is unique. This procedure
continues for the $n$ generations.
The method is hierarchical as only clusters that have the same number of
primary particles are used in each step.
We generated clusters composed of up to 4096 monomers with different $d_f$ and $k_f$.

Most of the clusters we examined were created with parameters\footnote{Henceforth,
we shall specify power-law clusters by the ordered pair ($d_{f},k_{f}$).}
characteristic of aggregates generated by Diffusion Limited Cluster
Aggregation - DLCA ($1.8, 1.3$)~\cite{Brasil00}
or Reaction Limited Cluster Aggregation - RLCA ($2.05, 0.94$) \cite{Lattuada03a}
The agglomeration mechanism for both groups is diffusion, the
difference arising from the cluster-monomer sticking probability:
it is unity for DLCA clusters and $10^{-3}$ for RLCA~\cite{Lattuada04}.
Note that Ref.~\cite{Lattuada03a} uses (1.85, 1.117) for DLCA-like clusters.

The morphology of the generated structures was analyzed to ensure
that they have the prescribed properties.
A double logarithmic plot of the number of monomers
versus the corresponding aggregate radius of gyration
[Eq.~(\ref{eq:r_gyr_definition})]
for clusters composed of $N = 2^n, n = 8 - 12$ monomers
and fixed ($d_f, k_f$) confirmed that the aggregates satisfy exactly
the scaling law.
The radius of gyration for $N$ equal, spherical monomers is calculated by
\begin{equation}
\label{eq:r_gyr_definition}
R_g^2 = \frac{1}{N} \, \s_{i=1}^N{ ({\bf r}_i -{\bf R}_{CM})^2} + \frac{3}{5} R_1^2,
\end{equation}
where ${\bf r}_i$ is the position of the $i$th monomer's center,
and the aggregate center of mass is ${\bf R}_{CM}=  1/N \s_{i=1}^N{\bf r}_i$.
Note that we included the additive term $3 R_1^2/5$ because we
are interested in the power-law dependence
even for small clusters; otherwise Eq.~(\ref{eq:r_gyr_definition}) evaluates to zero for a monomer.
This additional term may also be taken to be the square of the monomer radius~\cite{filippov, LorenzoPRE10}.
We chose $3 R_1^2/5$ because it is the radius of gyration of
a single 3d sphere of radius $R_1$~\cite{Lapuerta10, Sorensen97}.\footnote{We repeated our calculations
using $R_1^2$ as the additive term. We found minimal differences in the structural and dynamical
properties of the generated clusters, even though the numerical constants
in Eqs.~(\ref{eq:fitRmRg1},~\ref{eq:fitRmRg2}) differed slightly.}
Of course, for large $N$ the choice of the additional additive term is irrelevant.

An alternative, more precise, validation method
of the ``mimicking" algorithm is based on the two-point, orientationally-averaged
monomer-monomer correlation function $g(r)$. We calculated it
as follows: an ensemble of $M$ clusters composed of $N$ monomers
was generated, and all the pairwise $(i,j)$ Euclidean distances were
determined ($i,j = 1, \ldots N$).
The total number of distances is $N(N-1)$. The number of
particles $n_i(r)$ (equal to the number of distances)
within the interval  $[r-dr, r + dr]$ was recorded. We chose $dr = 0.1 R_1$, a value
we found to give reasonably smooth results~\cite{Lattuada03b}. The orientationally averaged,
spherically symmetric pair
correlation function is~\cite{filippov-review, Lattuada03b}
\beq
g(r) = \frac{1}{M} \sum_{i=1}^M \frac{n_{i}(r)}{4\pi r^{2} dr N},
\label{eq:NumericalPairCorrelation}
\eeq
with the normalization condition
\beq
N - 1 = 4 \pi \int_0^{\infty} \,d \, r \, r^2 g(r).
\label{eq:NumberParticlesPairCorr}
\eeq
The physical interpretation of $g(r)$
is that it gives the probability (per unit volume)
of finding a monomer at distance $r$ from an arbitrarily chosen
monomer~\cite{Heinson2012}.
Note that the pair correlation function, defined with respect to an arbitrarily
chosen monomer, is distinct
from the radial (mass) distribution function $\rho(r)$ which gives
the cluster (mass) distribution with respect to its center of mass.

\begin{figure}[htb]
\begin{centering}
\includegraphics[width=0.45\columnwidth, height=0.4\columnwidth]{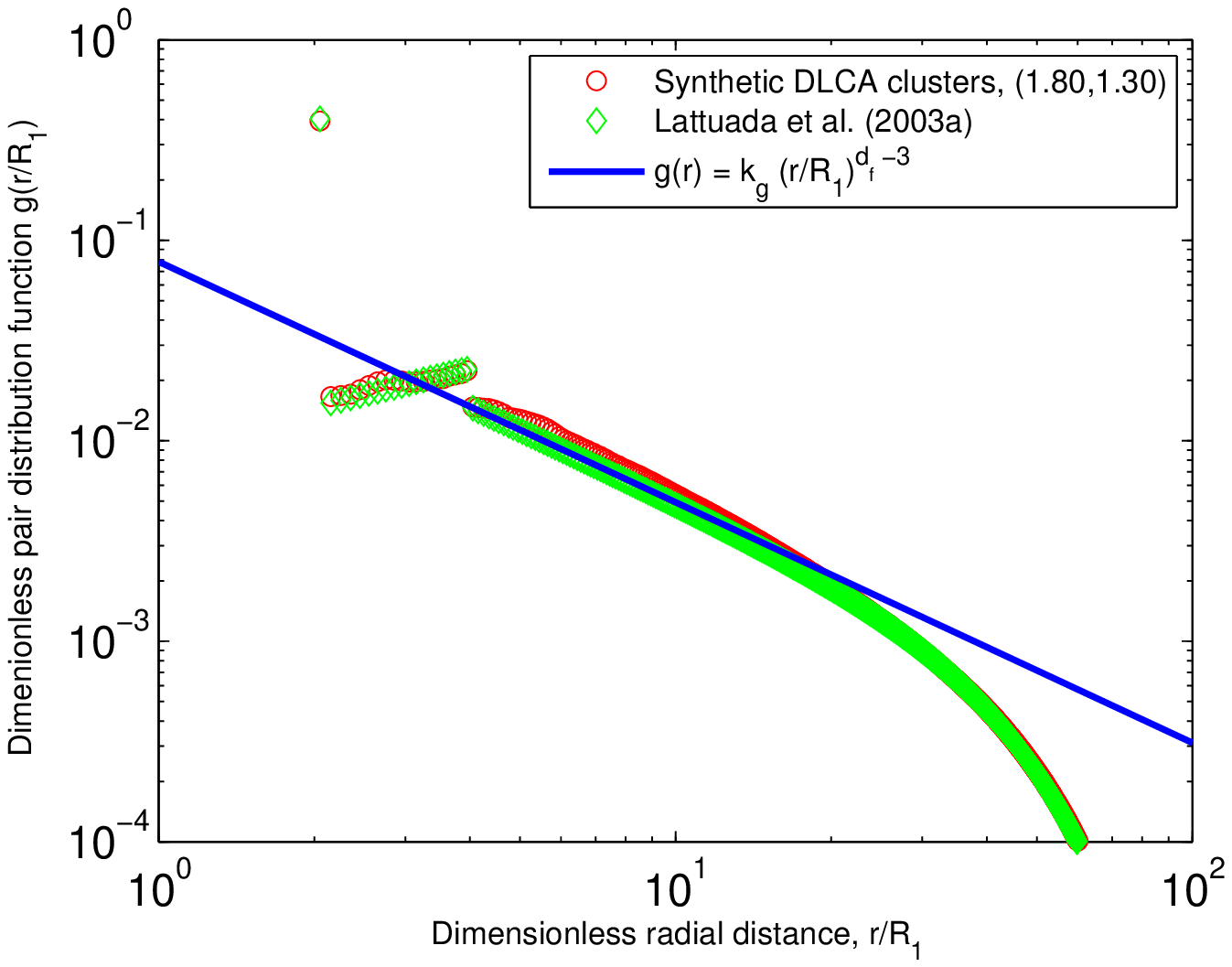}
\includegraphics[width=0.45\columnwidth, height=0.4\columnwidth]{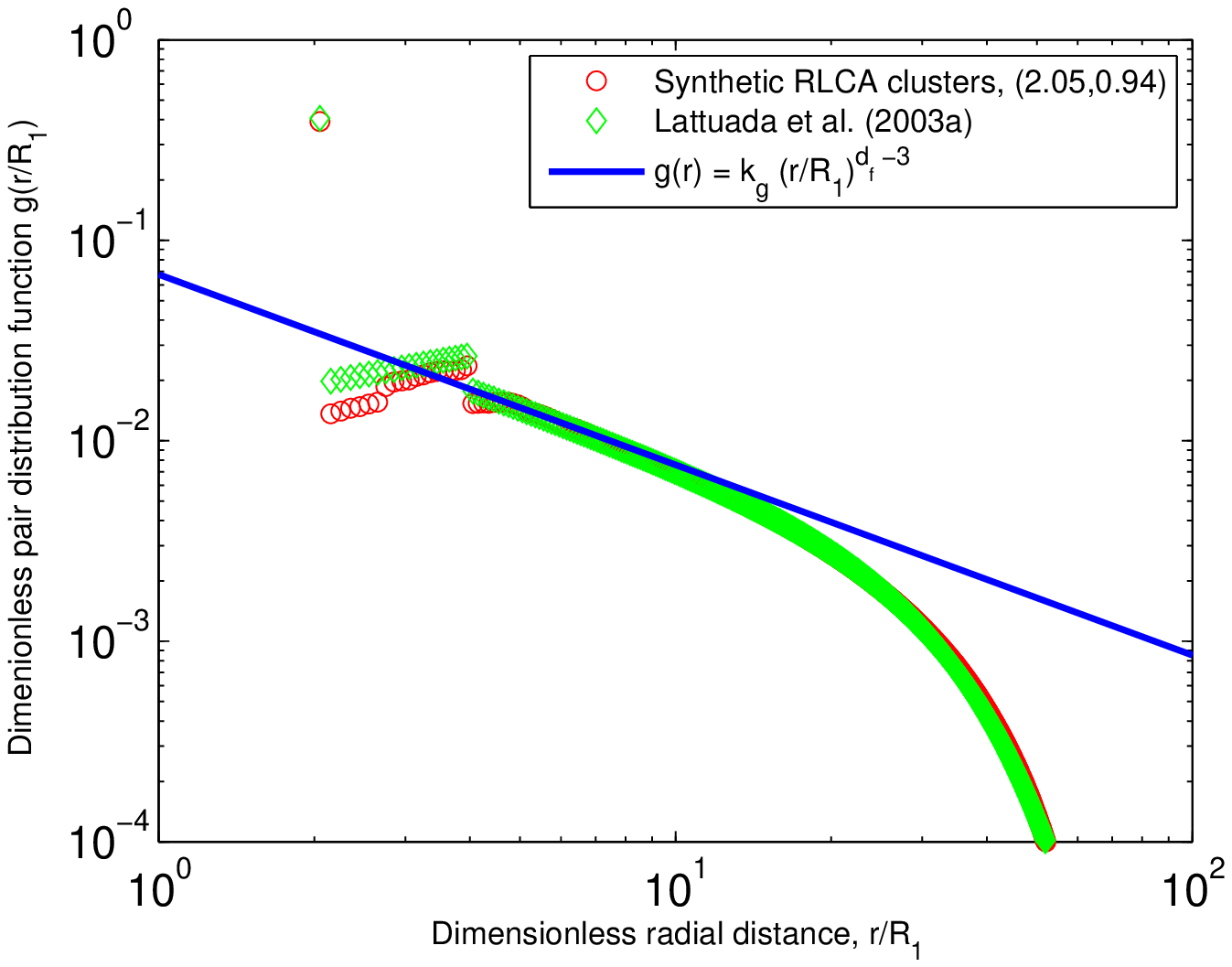}
\caption{Normalized, dimensionless, spherically symmetric two-point correlation function
ensemble-averaged over 2000 clusters with $512$ monomers. Left: DLCA clusters, $k_g = 0.0782$;
Right: RLCA clusters, $k_g = 0.0675$.}
\label{fig:CorrelationFunction}
\end{centering}
\end{figure}

An analytic expression for $g(r)$ is highly desirable as structural and dynamical
aggregate properties may be expressed in terms of it. The expected functional form is
\beq
g(r) = \frac{A}{R_1^{d_f}} \, r^{d_{f}-3} h \Big ( \frac{r}{\xi} \Big ),
\label{eq:PairCorrelationFunction}
\eeq
where $A$ is a constant. The algebraic decay arises from the scaling behaviour, and
the cut-off function $h(r/\xi)$ models finite-size effects. The correlation length $\xi$ is a measure
of the cluster's diffuse interface, the interface ``roughness".
The cut-off function is usually taken to be a stretched exponential,
\beq
h \Big ( \frac{r}{\xi} \Big ) = \exp \Big [ - \Big ( \frac{r}{\xi} \Big )^{\gamma} \Big ],
\label{eq:CutOffFunction}
\eeq
the stretching exponent $\gamma$ at most weakly dependent on the agglomeration mechanism.
As values are given $\gamma = 2.02$~\cite{Heinson2012}
or $2.20$~\cite{Lattuada03b} for  DLCA, and $2.16$ for RLCA clusters~\cite{Lattuada03b}.
The normalized, dimensionless pair correlation function $R_1^3 \, g(r/R_1)$ averaged over $2000$
aggregates consisting of $512$ monomers is plotted in Fig.~\ref{fig:CorrelationFunction}:
the left subfigure refers to DLCA aggregates, the right to RLCA aggregates.

The most accurate expression, so far, for $g(r)$ has been
proposed by Lattuada et al.~\cite{Lattuada03a}
who appreciated the importance of small-scale structure by identifying specific functional forms at
the first and second shells ($r=2 R_1$ and $2 R_1 < r < 4 R_1$).
The figure compares our results (``Synthetic clusters") to theirs.
The agreement is very good suggesting that the synthetic clusters exhibit
the expected power-law decay with the specified fractal exponent
($\df = 1.80$, left, and $\df = 2.05$, right)
over approximately one decade  (compare to the pure scaling-law expression, solid line).
The constant $k_g$ in the pure algebraic-decay expression was evaluated as suggested
in Ref.~\cite{Lattuada03a} (their constant $c$).

\subsection{Scaling law}
\label{sec: Scaling law}

The scaling law Eq.~(\ref{eq:scaling}) may also be re-written in terms of other
characteristic length scales, like the outer radius $\Rout$ and the
geometric radius $\Rgeo$.
The outer radius is defined as half the maximum distance between any two monomers
in the aggregate, whereas the geometric radius is the radius of the smallest sphere
encompassing the aggregate, centered at its center of mass (the smallest convex envelope of the
aggregate).\footnote{As the distance between monomers
is calculated with respect to the monomers center of mass, the monomer radius $R_1$
has been added to the calculation of both length scales to ensure the correct
single-monomer limit.}  Literature values for the ratio $\Rout / \Rg$, a ratio that can be used to determine
the radius of gyration from TEM images~\cite{Koylu95a}, vary by about 20\% for DLCA clusters,
being in the range [$1.45 - 1.65$]. We analyzed ensembles of 5000 clusters
consisting of up to 300 monomers to find that the ratio falls in the
range [$1.625 - 1.68$] (DLCA) and [$1.625 - 1.69$] (RLCA), cf. Fig.~S1
(Supplementary Material), the ratio depending weakly on $N$.
For approximately $100$ monomers the ratio evaluates to $\sim 1.675$ (DLCA) and $1.685$ (RLCA).
The ratio $\Rout / \Rgeo$ was determined to be $~0.915$
for both DLCA and RLCA synthetic clusters, largely
independent of the number of monomers, cf. Fig.~S2 (Supplementary Material).

We found, by performing linear fits on appropriate $\log N -\log R$ plots, that the
fractal dimension does not depend on the chosen geometric length scale, whereas
the prefactor does. In the case of the outer radius the prefactor is related to the average
cluster packing fraction $\phi$~\cite{Heinson2012}, while for the the geometric radius it becomes
the inverse of the volume filling factor $f$~\cite{Naumann}.
The fractal-like scaling law remains valid even if
expressed in terms of the hydrodynamic radius, as shown in Section~\ref{sec:MobilityScalingLaw};
however, the exponent, referred to as the mass-mobility exponent, differs
from the fractal exponent used in Eq.~(\ref{eq:scaling}).

It is important to note, as inspection of Fig.~\ref{fig:CorrelationFunction} shows,
that the aggregates considered herein are self-similar over a limited range of
monomer-monomer distances.
In particular, smaller clusters are not self similar and larger clusters have a diffuse interface.
Nevertheless, the fractal-like scaling law is valid for a number of choices
of the characteristic length scale, be it $R_g$, as in Eq.~(\ref{eq:scaling}),
$\Rout$, or $\Rgeo$ (or even the hydrodynamic radius,
cf. Sec.~\ref{sec:MobilityScalingLaw}); the validity of the scaling law
for the synthetic clusters is reflected in referring to them as power-law or fractal-like.
We use the scaling law irrespective of whether the aggregate has a real scale-invariant (self-similar)
morphology. This implies that we approximate an aggregate by an aggregate with a sharp
interface [$\gamma \to \infty$, see Eq.~(\ref{eq:CutOffFunction})] for which
the scaling law holds with respect to a well defined, outer length scale.

\section{Fractal dimension ($d_{f}$) and prefactor ($k_{f}$)}
\subsection{Small-scale structure}

The complex, and intricate, interdependence of $N$, $d_{f}$ and $k_{f}$, and the resulting
changes in the small- and large-scale structural properties of the aggregates, were investigated by
examining ensembles of 5000 aggregates. The parameter choices
and the calculated structural parameters are summarized in Table~\ref{table:Anisotropy}.
Different cluster ensembles are grouped according to the parameter that is investigated (in bold):
number of monomers (top group), fractal dimension (middle group), and fractal prefactor
(last group).
Note that $64$-monomer clusters defined by ($1.9, 1.3$) and ($1.8$, $1.6$) have identical
radii of gyration.
%

\begin{figure}[htb]
\begin{centering}
\includegraphics[width=0.65\columnwidth, height=0.5\columnwidth]{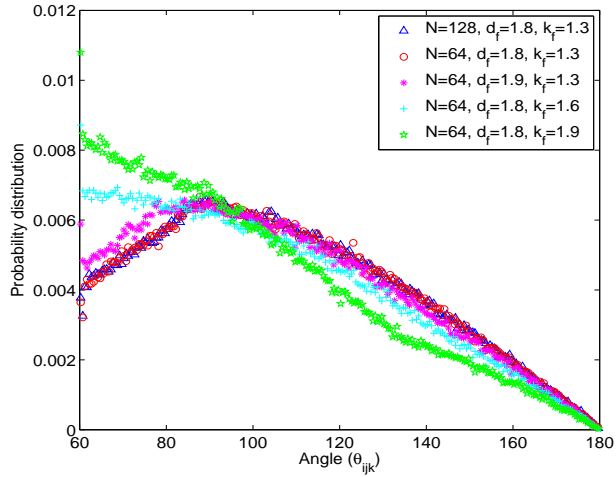}
\caption{Ensemble-averaged probability distributions of three-monomer cluster angles.}
\label{fig:Angle}
\end{centering}
\end{figure}

An indicator of a cluster's small-scale structure is the probability distribution of
the angles formed by three monomers. The angles are specified by two intersecting lines passing though
the center of mass of a central monomer $i$ and two $j,k$ monomers touching it.
For every monomer $i$ we calculated the
number of its neighbours $k$, to which we associated $k(k-1)/2$ angles
(possible pairwise combinations). We calculated the angles from the distance $d_{jk}$
of any two $(j,k)$ pairs via
\beq
\theta_{ijk} = 2 \sin^{-1} \Big ( \frac{d_{jk}}{4 R_1} \Big ).
\label{eq:thetaijk}
\eeq
\begin{table}[h]
\caption{Mean characteristic structural parameters: shape anisotropy $ \langle A_{13} \rangle$
and three-monomer angle $\langle \theta_{ijk} \rangle$.}
\label{table:Anisotropy}
\begin{center}
\begin{tabular}{cccccc} \hline \hline
& $N$ & ($d_{f}, k_{f}$) & $R_{g}/R_{1}$& $\langle A_{13} \rangle$ &
$\langle \theta_{ijk} \rangle$ \\ \hline 
1 & \textbf{512} & (1.8, 1.3) & 27.7 & 3.82 & 107.4 \\ 
2 & \textbf{256} & (1.8, 1.3) & 18.8 & 3.77 & 107.3 \\ 
3 & \textbf{128} & (1.8, 1.3) & 12.8 & 3.70 & 107.3 \\ 
4 & \textbf{64} & (1.8, 1.3) & 8.7 & 3.69 & 107.2 \\ 
4-bis & \textbf{32} & (1.8, 1.3) & 5.93 & 3.69 & 106.9 \\  \hline 
4 & 64 & (\textbf{1.8}, 1.3) & 8.7 & 3.69 & 107.2 \\ 
5 & 64 & (\textbf{1.9}, 1.3) & 7.8 & 3.23 & 105.4 \\ 
6 & 64 & (\textbf{2.0}, 1.3) & 7.0 & 2.90 & 103.8 \\ 
7 & 64 & (\textbf{2.1}, 1.3) & 6.4 & 2.60 & 102.4 \\ \hline 
4 & 64 & (1.8, \textbf{1.3}) & 8.7 & 3.69 & 107.2 \\ 
8 & 64 & (1.8, \textbf{1.6}) & 7.8 & 3.52 & 102.6 \\ 
9 & 64 & (1.8, \textbf{1.9}) & 7.1 & 3.40 & 98.8 \\ 
10 & 64 & (1.8, \textbf{2.2}) & 6.5 & 3.30 & 92.6 \\ \hline \hline 
\end{tabular}
\end{center}
\end{table}
Figure~\ref{fig:Angle} presents the resulting distributions of three-monomer angles.
The angles vary from $60^\circ$, the minimum possible angle for three
touching equal-sized spheres their centers forming an equilateral triangle, and
$180^\circ$, a locally straight chain configuration. Angles less than $60^\circ$ would
imply monomer overlapping or ``necking''.
The mean values reported in Table~\ref{table:Anisotropy} (upper group) are in reasonable
agreement with the previously reported value~\cite{Lattuada03b} for trimer distributions of
($1.85, 1.117$) DLCA aggregates, $103.57^\circ$.
The distributions are independent of $N$, while they depend weakly on
$d_{f}$ and strongly on $k_{f}$. We note that as the prefactor increases the number of
small angles ($\leq 80^\circ$) increases, suggesting that the prefactor is
an indicator of local structure; for fixed $d_f$ as the prefactor increases
the cluster becomes more locally compact ($\langle \theta_{ijk} \rangle$ decreases).
This observation is also supported
by the mean angles presented in Table~\ref{table:Anisotropy},
and the quantitative comparison presented in Table~S1 (Supplementary Material).

An alternative indicator of local compactness
is the mean number of nearest neighbors (number of
touching monomers), or coordination number $c_N$,
defined as the average number of contacts
a monomer has within an aggregate~\cite{friedlander-book}.
The  cluster coordination number not only provides information on
the openness of an aggregate and its compactness, but it
is a factor that influences monomer hydrodynamic shielding within an aggregate~\cite{sonntag}.
Reference~\cite{LorenzoPRE10} used the coordination number
as an indicator of cluster compactness, albeit for clusters
generated by a completely different, physically-based agglomeration mechanism.
For the synthetic fractals analyzed herein, i.e., generated by the cluster-cluster
aggregation algorithm, Gastaldi and Vanni~\cite{Gastaldi} argued that the
the coordination number is
\beq
c_N = 2 \frac{N-1}{N}.
\label{eq:CoordinationNumberVanni}
\eeq
We checked this expression for clusters composed
of $N=16,32,64,128$ with different ($\df, \kf$): we found it to be
very accurate.
Figures~S3 and S4 (Supplementary Material) present probability distributions of the number of nearest neighbours
for cluster ensembles specified by $N=64, 128$, $\df = 1.5, 1.8$ and $\kf = 1.3, 2.2$.
As the prefactor increases the distribution function broadens, but the
coordination number is only a function of the number of monomers, as suggested
by Eq.~(\ref{eq:CoordinationNumberVanni}).
%


\subsection{Large-scale structure}

One recently used large-scale indicator is the cluster shape anisotropy $A_{13}$,
a measure of cluster stringiness; for example, as $A_{13}$ increases the aggregate
becomes more cigar-like. The shape anisotropy is calculated from the
principal radii of gyration $R_i$ ($i=1,2,3$) by diagonalizing
the aggregate inertia tensor~\cite{Heinson2010}. Accordingly,
the radius of gyration may be written as~\cite{Heinson2012}
\beq
R_{g}^{2}=\frac{1}{2}(R_{1}^{2}+R_{2}^{2}+R_{3}^{2}), \quad R_{1}\geq R_{2}\geq R_{3},
\eeq
and the shape anisotropy $A_{13}$ is defined by
\beq
A_{13}=\frac{R_{1}^{2}}{R_{3}^{2}}.
\eeq

\begin{figure}[htb]
\begin{centering}
\includegraphics[width=0.65\columnwidth, height=0.5\columnwidth]{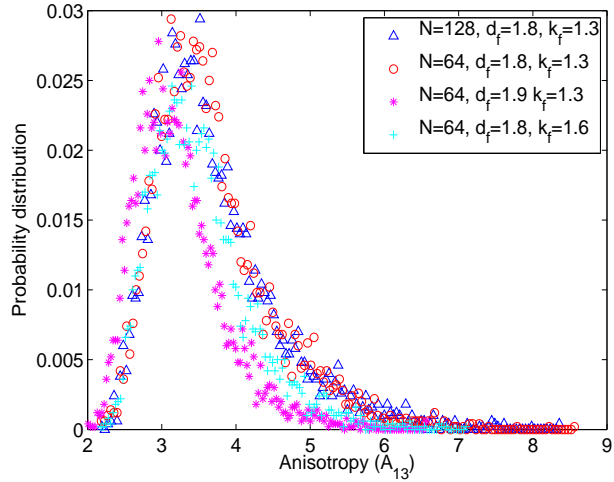}
\caption{Probability distribution of shape anisotropy $A_{13}$.}
\label{fig:A13}
\end{centering}
\end{figure}

Figure~\ref{fig:A13} presents probability distributions of shape anisotropies for different ensembles of fractals.
We observe that $A_{13}$ depends strongly on $d_{f}$, weakly on $k_{f}$,
and is independent of $N$. These observations are confirmed by the mean $\langle A_{13} \rangle$
reported in Table~\ref{table:Anisotropy}. We remark that
anisotropies extend over a large range of values, even for the same ($d_{f}, k_{f}$).
As expected, our results agree with Thouy and Jullien~\cite{Thouy2}, who concluded
that (for fixed $k_f$) shape anisotropy is independent of $N$ and dependent on $d_f$.

It is worthwhile to compare our results to Heinson et al.~\cite{Heinson2012, Heinson2010} who argued that
shape anisotropy affects the prefactor, rendering $k_f$ a shape indicator.
Our results are in agreement on the importance of the prefactor as descriptor
of an aggregate morphology: in fact, we identify synthetic clusters
by the ordered pair ($d_f, k_f$), and the calculated mean shape anisotropy for DLCA fractals
(Table~\ref{table:Anisotropy}, upper group)
is in reasonable agreement with their reported value
$\langle A_{13} \rangle = 3.86$~\cite{Heinson2012}.
Moreover,  we find that, for \textit{fixed} $d_f$, the prefactor $k_f$ depends on the mean anisotropy, albeit weakly.
Since the synthetic clusters are generated by specifying the fractal prefactor,
the argument that $\langle A_{13} \rangle$, via the prefactor,
describes aggregate structure at large length scales,
may be inverted, emphasizing the importance of $k_f$ to determine $\langle A_{13} \rangle$.
For fixed fractal dimension,
the two approaches are equivalent, i.e., if $\langle A_{13} \rangle$ increases $k_f$
decreases and vice versa, suggesting that local structure has an effect on large-scale
structure and vice versa.
If, however, $d_f$ is allowed to change we find that
the effect of its change on the shape anisotropy distribution (and, specifically, on its mean) is more
important than the effect of a change of the prefactor.

These observations on the effect of structural parameters
on aggregate morphology are summarized in Table~\ref{table:Anisotropy}.
A comparison of clusters pertaining to the middle group
shows the effect of $d_f$, whereas a comparison of
the lower group shows the effect of $k_f$.
They indicate that changes of the fractal dimension
produce larger changes of the mean shape anisotropy, and changes of the prefactor larger changes of the
mean three-monomer angles.
The comparisons are rendered quantitative in Table~S1 (Supplementary Material), where
the effect of variations of the scaling-law parameters $\df, \kf$ and
the number of monomers $N$ on mean characteristic cluster
structural parameters is presented as appropriate percentage changes.

Hence, in general, the fractal dimension is an
indicator of the overall aggregate shape (large-scale aggregate morphology), while the prefactor
becomes an indicator of local structure (small-scale morphology).
Mean shape anisotropy (an indicator of the aggregate shape) is important
as it affects the value of the prefactor.

\section{Aggregate structure and hydrodynamic radius}
\subsection{Dependence of the hydrodynamic radius on the radius of gyration}

\begin{figure}[thb]
\begin{center}
\includegraphics[width=0.65\columnwidth, height=0.5\columnwidth]{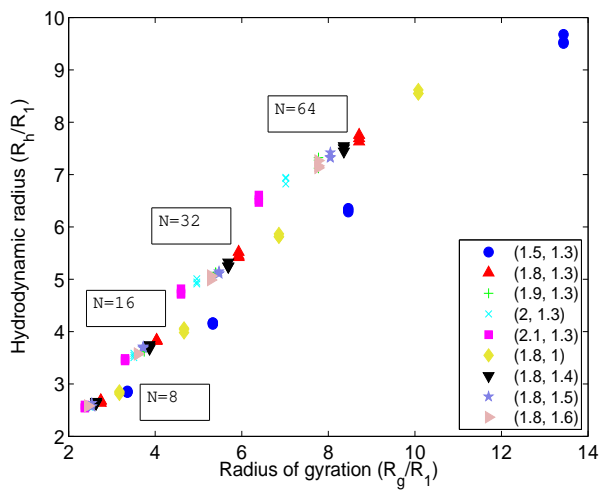}
\caption{Hydrodynamic radius as a function of the radius of gyration.}
\label{fig:RmRg}
\end{center}
\end{figure}

We calculated the hydrodynamic radii of
clusters composed of $8, 16, 32, 64$ primary particles with
($d_{f}, k_f$) in the ranges ([$1.5, 2.1$], [$1, 1.6$]). We simulated
three realizations of nine different
($d_f,k_f$) pairs for each $N$.
Calculated hydrodynamic radii are plotted against the corresponding radii of
gyration in Fig.~\ref{fig:RmRg}.
Each symbol represents a single aggregate.

A striking feature of the figure is that the calculated $R_h$
cluster according to the number of monomers, suggesting a
linear ($R_h$, $R_g$) relationship for a given monomer number.
Accordingly, for fixed $N$, we fit linearly the data to
\beq
\frac{R_{h}}{R_1} = m(N) \frac{R_{g}}{R_1} + b(N),
\label{eq:general}
\eeq
the slope $m$ and the $y$-intercept $b$ being functions of $N$. The resulting four
$m(N)$ and $b(N)$ are averaged to obtain the final empirical fit. In fact,
we performed two different fits: one with the independent variable
being the equivalent volume radius ($R_{eq}= R_1 N^{1/3}$)
\beq
\frac{R_{h}}{R_1} = 0.248 \, \Big ( 2 - N^{-1/3} \Big ) \, \frac{R_{g}}{R_1} + 0.69 N^{0.415},
\label{eq:fitRmRg1}
\eeq
and one with $\ln(2N)$, a dependence suggested by the ($R_h, R_g$) relationship
for straight chains (see, for example, Ref.~\cite{LorenzoJCIS11}),
\beq
\frac{R_{h}}{R_1} = 0.548 \, \Big [ 1 - \frac{1}{\ln(2N)} \Big ] \, \frac{R_{g}}{R_1} + 0.73 N^{0.40}.
\label{eq:fitRmRg2}
\eeq

\begin{figure}[thb]
\begin{centering}
\includegraphics[width=0.45\columnwidth,height=0.4\columnwidth]{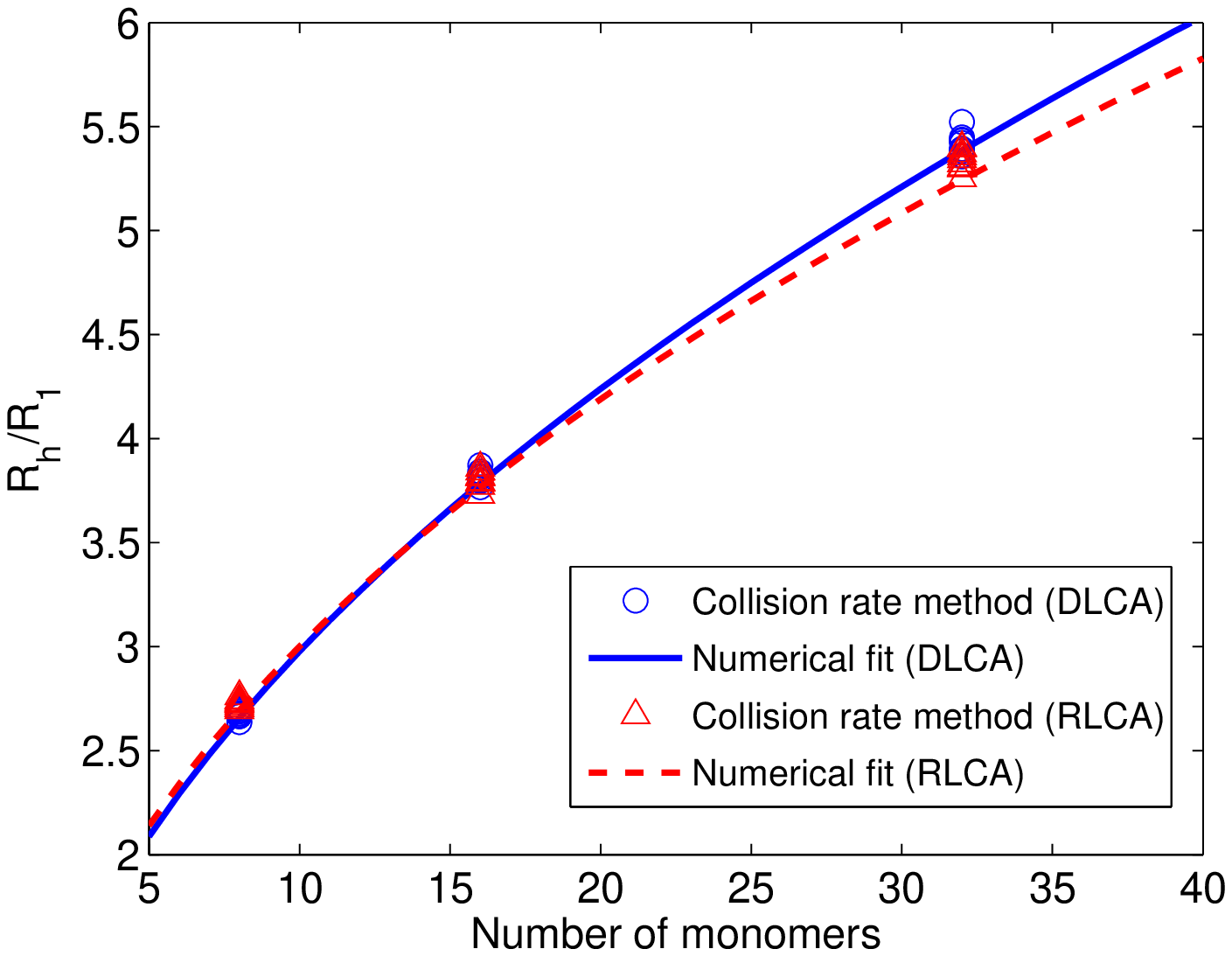}
\includegraphics[width=0.45\columnwidth,height=0.4\columnwidth]{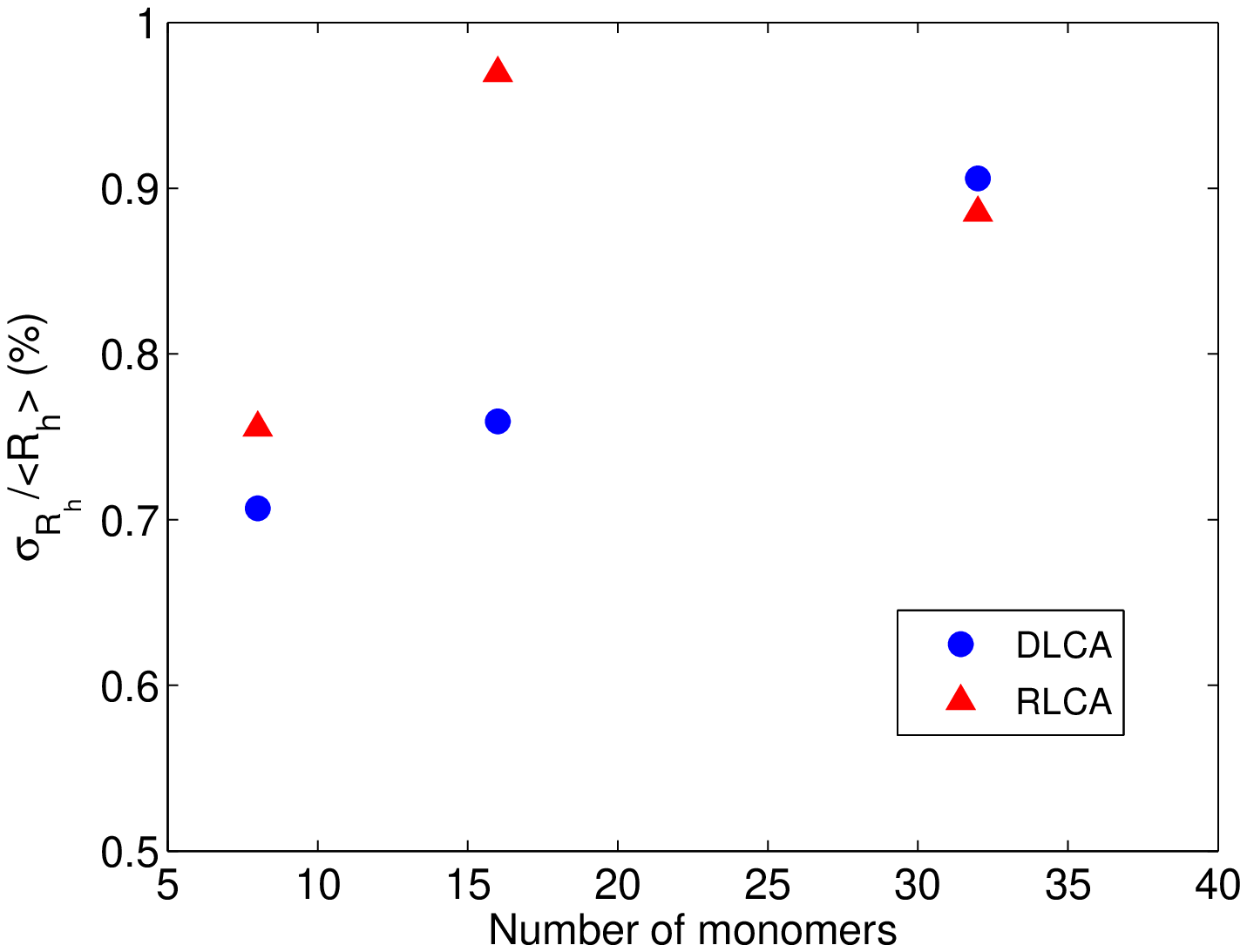}
\caption{Left: Calculated hydrodynamic radii of 10 DLCA and 10 RLCA clusters (symbols). Comparison with
numerical-fit predictions (lines); Right: Variability of DLCA and RLCA hydrodynamic radii: percentage ratio
of the hydrodynamic-radius standard deviation to the mean hydrodynamic radius.}
\label{fig:HydrodynamicRadiusVariability}
\end{centering}
\end{figure}

Since the numerical fits were obtained from three different ($N, \df, \kf$) realizations
[corresponding, nevertheless, to $27 \times 4 = 108$ ($N, R_g$) realizations], we estimated the variability
of the hydrodynamic radius for two ($N, \df, \kf$) choices. We calculated the hydrodynamic
radius of 10 DLCA and 10 RLCA clusters to obtain the
mean hydrodynamic radius, $\langle R_h \rangle$, and an estimate
of the hydrodynamic-radius variability, herein chosen to be the ratio of the
hydrodynamic radius standard deviation to the mean hydrodynamic radius,
$\sigma_{R_h} / \langle R_h \rangle$ (expressed as a percentage).
Results are shown in Fig.~\ref{fig:HydrodynamicRadiusVariability}. The left subfigure
presents the calculated hydrodynamic radii for each
cluster realization and the numerical fit: the agreement is very good.
Note that the hydrodynamic-radius variability is so small that error bars would not have
been visible. The right subfigure presents the chosen measure of
the variability. It is important to note that the variability of $R_h$ is so small
that even a limited number of ($N, \df, \kf$) triplet realizations would cover a large range
of $R_h$ values, thereby justifying our choice to use a limited number of triplets.

Equations~(\ref{eq:fitRmRg1}, \ref{eq:fitRmRg2}) suggest
that neither the fractal dimension
nor the fractal prefactor are separately necessary to estimate $R_{h}$, as it may be
fitted solely on $N$, $R_{g}$ (and, of course, $R_1$).
The general dependence $R_{h} = f(N,k_{f},d_{f})$ may, thus, be simplified
via the implicit dependence on $k_f$ and $d_f$ through $R_g$, $R_{h} = f(N, \, R_{g}(N,k_{f},d_{f}))$.
This observation should be contrasted to most empirical fits in the literature
where the hydrodynamic radius is expressed in terms of $R_g, d_f$, and possibly $N$
see, for example, Refs.~\cite{Thajudeen,Naumann}. Moreover, Eqs.~(\ref{eq:fitRmRg1}, \ref{eq:fitRmRg2})
imply that $R_{h}$ may be calculated for a single cluster, if the monomer positions
are known (from simulations or experimental measurements),
since the independent variables do not depend on ensemble-averaged properties like $d_f$ and $k_f$.

\begin{figure}[thb]
\begin{centering}
\includegraphics[width=0.45\columnwidth,height=0.4\columnwidth]{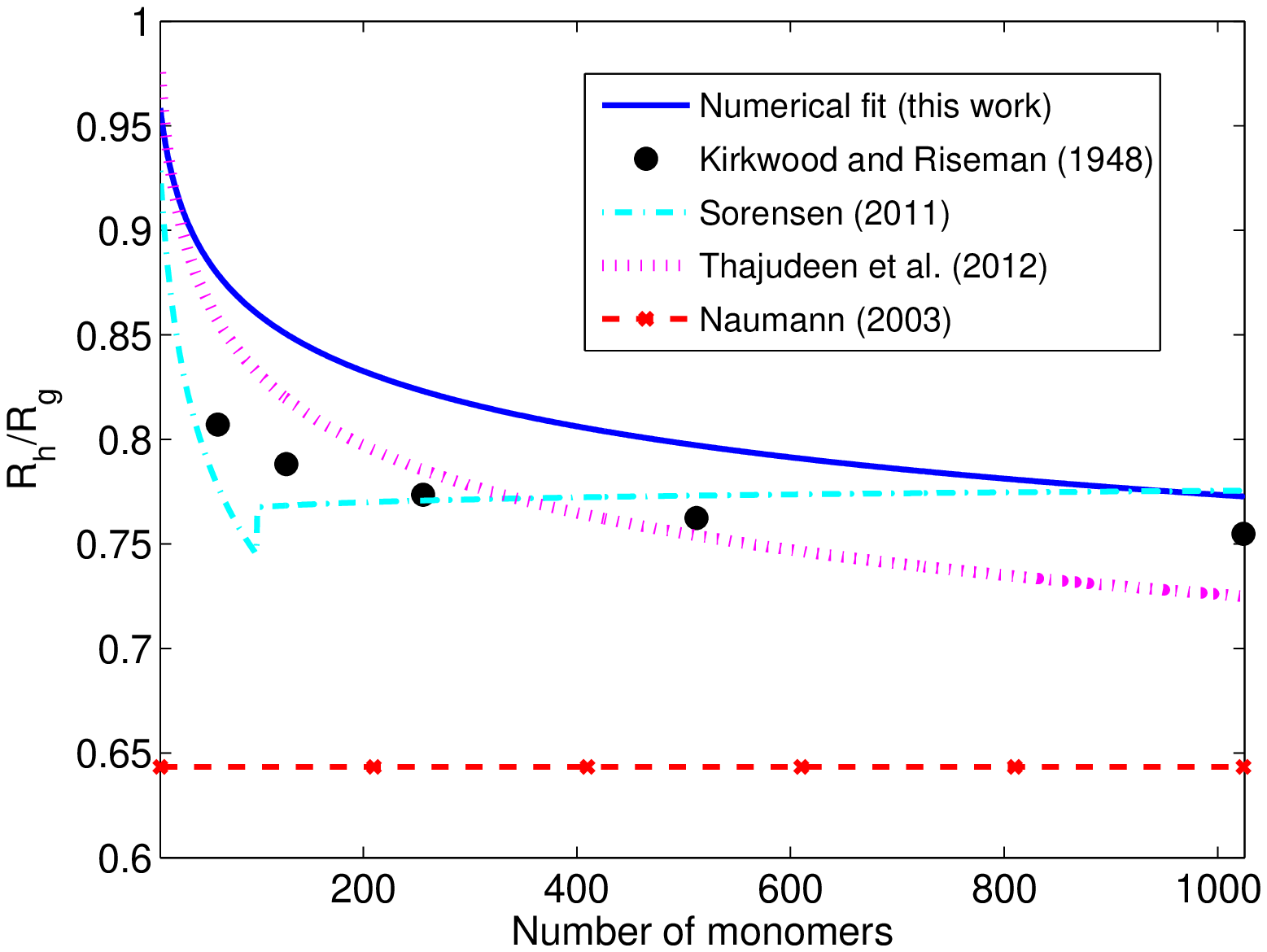}
\includegraphics[width=0.45\columnwidth,height=0.4\columnwidth]{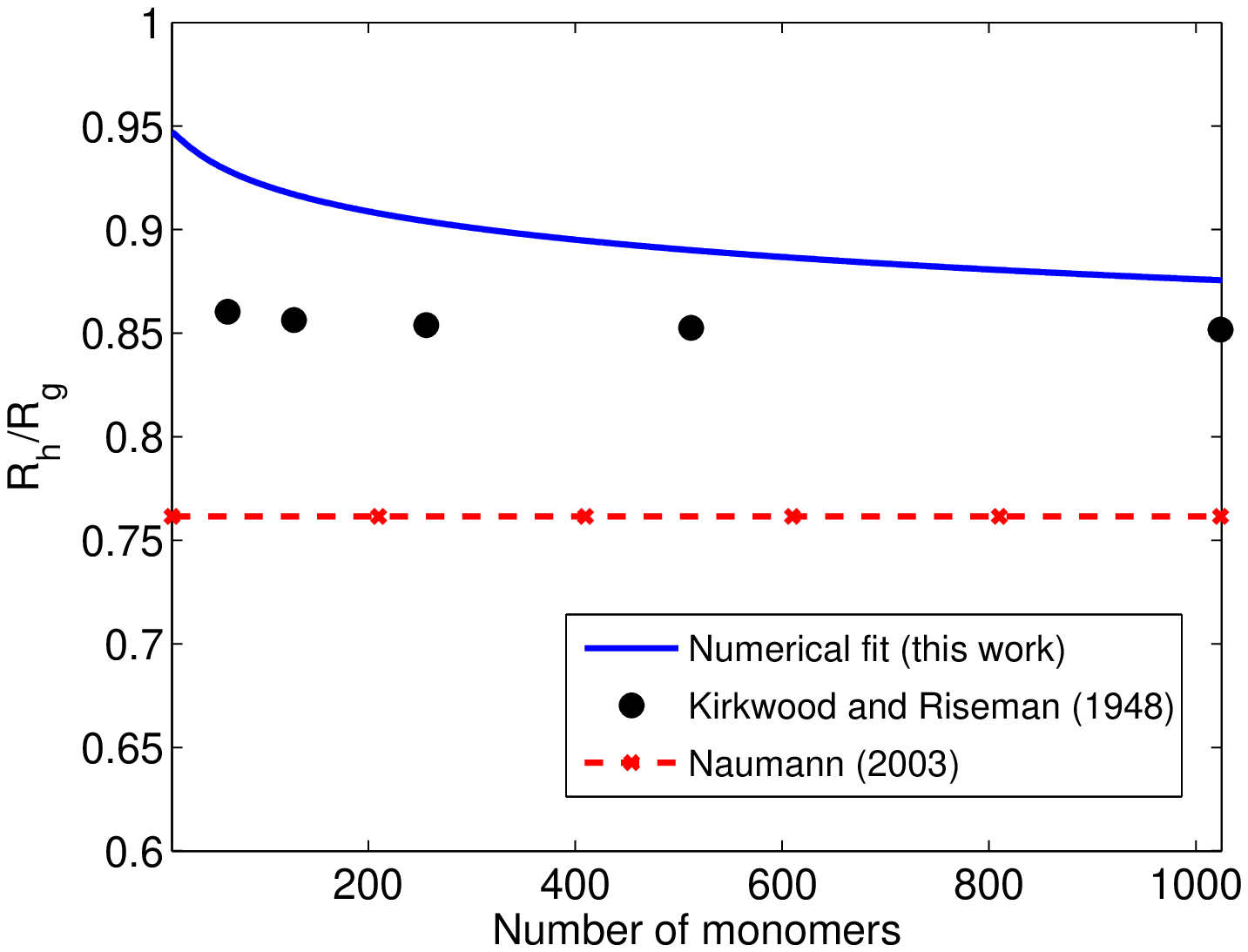}
\caption{Comparison of predicted ratios of the hydrodynamic to the radius of gyration. Left: DLCA clusters; Right: RLCA clusters.}
\label{fig:DLCARLCA}
\end{centering}
\end{figure}

Henceforth, we will use Eq.~(\ref{eq:fitRmRg1}) as the predicted hydrodynamic
radii $R_{h}$
are almost identical, irrespective of which equation is used.
Figure~\ref{fig:DLCARLCA} compares the numerically determined
ratio $R_{h}/R_{g}$  for up to $N=1000$ to  previously proposed
theoretical~\cite{KR}, semi-analytical~\cite{Naumann},
and numerical~\cite{Thajudeen,Sorensen_re} expressions. The left subfigure refers to DLCA clusters,
whereas the right subfigure to RLCA clusters. Note that as $N \rightarrow \infty$ the ratio tends to a constant
characteristic of the agglomeration mechanism.

Our results compare favorably to the purely geometric expression derived by Kirkwood
and Riseman~\cite{KR}, according to which
\beq
\frac{N R_1}{R_h} = 1 + \frac{1}{N} \sum_{j=1}^N \sum_{m=1, m \neq j}^N \frac{R_1}{|\mathbf{r}_j -\mathbf{r}_m|}.
\label{eq:KR}
\eeq
Predictions of Eq.~(\ref{eq:KR}) are lower than the empirical-fit
predictions for small aggregates, approaching the same limit as $N$ increases.
The comparison suggests that the Kirkwood-Riseman expression gives
a very good approximation to the hydrodynamic radius of both open and closed
structures, the difference increasing as the number of monomers
decreases (as expected since it is a large $N$ expression).

Our predictions are also compared to the recently suggested expression by Thajudeen et al.~\cite{Thajudeen},
\beq
R_s \approx R_h = \frac{R_g}{\alpha_1 (d_f) + \alpha_2 (d_f)},
\label{eq:HoganFit}
\eeq
where $R_s$ is the Smoluchowski radius, taken to be approximately equal to the hydrodynamic
radius (as in this work), and $\alpha_i$ ($i=1,2$) are quadratic functions of $d_f$.
Note that Eq.~(\ref{eq:HoganFit}) depends explicitly on $d_f$, and it
has six fitting parameters. Its range of validity is
$k_f = 1.3$ and $d_f$ in the range [$1.30 - 2.60$]; hence, the
calculation of the hydrodynamic radius of RLCA clusters
is beyond its region of validity.

The fit proposed by Naumann~\cite{Naumann} underestimates the ratio $R_{h}/R_{g}$.
A possible reason is that $R_{h}$ is expressed in terms of $\Rgeo$;
for our synthetic clusters $\Rgeo/R_{g}=1.83$ (DLCA) and $1.84$ (RLCA)
(see Section \ref{sec: Scaling law}) values different from
the analytical expression
$\Rgeo/R_{g}=[(d_f+2)/d_f]^{1/2}$ that evaluates to $1.45$ (DLCA) and $1.41$ (RLCA).
A larger ratio would result in larger $R_h/R_g$, closer to the collision-rate results.

Figure~\ref{fig:DLCARLCA} also compares our DLCA results to the expression proposed
in the recent review of the mobility of fractal
aggregates~\cite{Sorensen_re}. The suggested expression is a piece-wise
continuous function, the segments matching at $N=100$, but with
a crossover at  $N=74$.

Our results for the ratio $R_h/R_g$ are also compared to available
numerical calculations and experimental measurements
for specific cluster parameters in Table~\ref{table:comparison}.
They differ from Filippov's calculations~\cite{filippov} by less than $10\%$,
providing further support of the validity of our proposed expression.
Lattuada et al.~\cite{Lattuada03a} calculated of same ratio for fractals generated by a Monte Carlo
cluster-cluster aggregation method via the Kirkwood-Riseman method.
Again, the agreement is very good.
\begin{table}[htb]
\caption{Comparison of numerically determined ratio $R_{h}/R_{g}$ with literature values.}
\label{table:comparison}
\begin{center}
\begin{tabular}{ccccc} \hline \hline
& $N$ & ($d_{f}, k_{f}$) & $R_{h}/R_{g}$ & Eq.~(\ref{eq:fitRmRg1}) \\ \hline
Fillipov~\cite{filippov} & 100 & ($1.8, 1.3$) & 0.89 & 0.86 \\
& 100 & ($1.8, 2.3$) & 0.98 & 1.02 \\
& 100 & ($1.2, 2.5$) & 0.60& 0.66 \\ \hline
Lattuada & 1000 & ($1.85, 1.117$) DLCA & 0.77 & 0.78 \\
et al.~\cite{Lattuada03a} & 1000 & ($2.05, 0.94$) RLCA & 0.83 & 0.88 \\ \hline
Wang and & 1000 & ($1.75$, not specified) & 0.7 & 0.74 ($k_f = 1.3$) \\
Sorensen~\cite{Wang99} & 1000 & ($2.15$, not specified) & 0.97 & 1.02 ($k_f = 1.3$)\\
\hline \hline
\end{tabular}
\end{center}
\end{table}

\begin{figure}[htb]
\begin{centering}
\includegraphics[width=0.65\columnwidth,height=0.50\columnwidth]{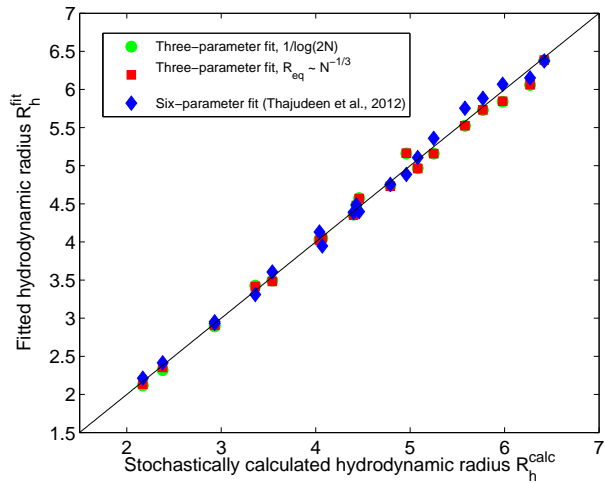}
\caption{Comparison of two numerical fits, Eqs. (\ref{eq:fitRmRg1}, \ref{eq:fitRmRg2}),
to calculated hydrodynamic radii, Ref.~\cite{Thajudeen}.}
\label{fig:Hogan}
\end{centering}
\end{figure}

As an additional confirmation of the accuracy
of our empirical expression we considered the $20$ structures
discussed in detail in Ref.~\cite{Thajudeen}, their Table~$1$. We
found, Fig.~\ref{fig:Hogan}, that predictions are in excellent agreement with the
six-parameter fit, the differences being at maximum $\pm 5$\%.
This result is not surprising as the two methodologies are
very similar: the collision-rate methodology obtains the hydrodynamic
radius from the solution of a diffusion equation, whereas the methodology
used in Ref.~\cite{Thajudeen} is based on averages of
particle-trajectory properties calculated from
the corresponding Langevin equations.
%
%

The proposed relationship between the hydrodynamic radius and the radius of gyration
may be easily converted into an expression relating the hydrodynamic radius
to the outer radius, a quantity sometimes easier to determine experimentally
than the radius of gyration (see Section~\ref{sec: Scaling law} for an estimate
of $\Rout/\Rg$).

\subsection{Power-law aggregates generated by Langevin dynamics}

An important feature of the proposed fit is that its application
does not require explicitly the cluster statistical
properties (ensemble averaged) $d_f$ and $k_f$. The fitting parameters depend only on morphological
(geometric) properties, as does the Kirkwood-Riseman expression.
Consequently, it may be used to calculate the hydrodynamic radius of clusters
given only their geometry.

A specific example of the usefulness of our numerical fit
is provided by considering the power-law aggregates generated in Ref.~\cite{LorenzoPRE10}.
These aggregates were generated by solving the Langevin equations of motion of a set
of monomers interacting via a central potential in a quiescent fluid.
The easily determined, instantaneous properties of these structures are geometric: the radius of gyration and
the number of primary particles.
The proposed fit provides an efficient formula to estimate
the diffusion coefficient of aggregates as they are being  formed, and thus to determine
aggregate formation without relying on the so-called ``free draining" approximation
for the hydrodynamic shielding of monomers within a cluster.
%

\begin{figure}[htb]
\begin{centering}
\includegraphics[width=0.65\columnwidth,height=0.50\columnwidth]{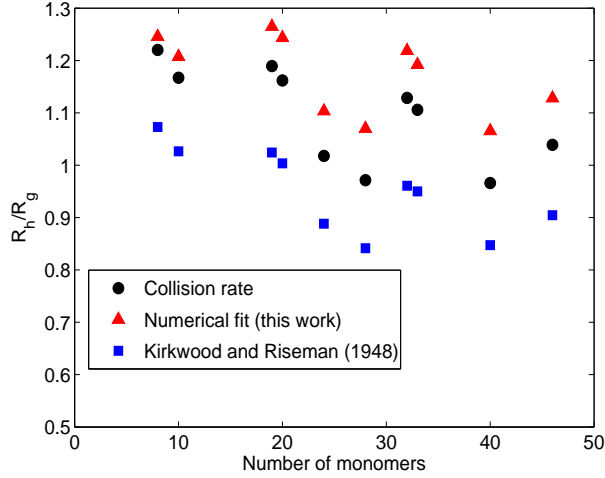}
\caption{Ratio of the hydrodynamic radius to the radius of gyration for power-law aggregates
generated via Langevin-dynamics simulations: Collision-rate calculation, numerical fit, and Kirkwood-Riseman predictions.}
\label{fig:LangevinAggregates}
\end{centering}
\end{figure}

We calculated the hydrodynamic radii by the collision-rate methodology,
and we compare them to predictions of the Kirkwood-Riseman theory and
the proposed expression Eq.~(\ref{eq:fitRmRg1}) in Fig.~\ref{fig:LangevinAggregates}.
The very good agreement (maximum deviation $10$\%) indicates that our fit
reproduces the hydrodynamic radii even for power-law aggregates generated by other methods.
We note that due to the choice of a
spherically symmetric monomer-monomer interaction potential, the
Langevin-dynamics generated power-law aggregates were locally compact
(large clusters at late time $\kf = 3.65$), and on larger
scales tubular and elongated (large clusters at late time $\df = 1.56$). Thus,
the comparison provides a rather
stringent test of the proposed expression.
As for the small-cluster comparison shown in Fig.~\ref{fig:DLCARLCA}, the Kirkwood-Riseman expression provides
a good approximation to the hydrodynamic radii, albeit slightly under-predicting them.

\subsection{Mobility scaling law and dynamic shape factor}
\label{sec:MobilityScalingLaw}

It has been argued that the scaling law remains valid even when
the characteristic length scale is chosen to be the mobility radius
(equal to the hydrodynamic radius in the continuum regime). The corresponding
scaling law is
\beq
N=k_{m} \Big( \frac{R_{h}}{R_{1}} \Big)^{d_{m}},
\label{eq:MassScaling}
\eeq
where $d_m$ is the mass-mobility exponent. We fitted the hydrodynamic radius calculated for DLCA and RLCA
aggregates to the number of primary particles (log-log fit) to obtain
\begin{subequations}
\beq
N=1.17 \big(\frac{R_{h}}{R_{1}}\big)^{1.97}; \quad \quad \textrm{DLCA} \ (1.8, 1.3),
\eeq
\beq
N=0.92 \big(\frac{R_{h}}{R_{1}}\big)^{2.14}; \quad \quad \textrm{RLCA} \ (2.05, 0.94).
\eeq
\end{subequations}
Thus, even though for different geometric radii the fractal dimension remains the same,
when a dynamic length scale is used the fractal dimension changes~\cite{Sorensen_re,Gmachowski02}. Of course, the
corresponding fractal prefactors
change.

The mass mobility exponent may be related to the fractal dimension by combining Eqs.~(\ref{eq:scaling}, \ref{eq:MassScaling})
to obtain
\beq
d_m = \df \, \frac{\log R_g}{\log R_h} \, \Big [ 1 + \frac{\log (\kf/k_m)}{\df \log R_g} \Big ].
\label{eq:RelateExponents}
\eeq
We found that use of the proposed expression for the hydrodynamic radius in Eq.~(\ref{eq:RelateExponents})
reproduces $d_m$ to within less than 0.5\%.
Moreover, the first term on the right-hand-side approximates $d_m$ to within $5$\% (DLCA) and
$1.5$\% (RLCA).
Note, however, that the difference between the calculated fractal dimension and the mass
mobility dimension is $10$\% (DLCA) and $5$\% (RLCA).

\begin{figure}[htb]
\begin{centering}
\includegraphics[width=0.65\columnwidth,height=0.50\columnwidth]{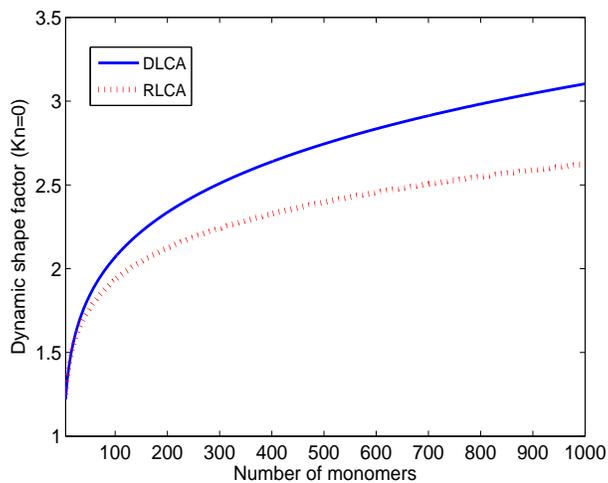}
\caption{Calculated dynamic shape factor of DLCA and RLCA clusters.}
\label{fig:DynamicShapeFactor}
\end{centering}
\end{figure}

Lastly, the empirical fit may be used to obtain the dynamic shape factor
$\chi_N$, a correction factor used
to account for the effect of the aggregate shape on its motion. It becomes~\cite{LorenzoJCIS11}
\beq
\chi_N = \frac{R_h}{R_1} \, N^{-1/3}.
\label{eq:DynamicShapeFactor}
\eeq
Figure~\ref{fig:DynamicShapeFactor} presents the calculated values for both
DLCA and RLCA clusters as a function of monomer number.
The dynamic shape factor may also be used to define the cluster effective density, the density
of a fictitious spherical particle of radius the hydrodynamic radius and of the same mass as the initial
irregularly shaped aggregate. It is defined by
\beq
\rho_{\textrm{eff}}=\rho_1 N \Big(\frac{R_1}{R_h}\Big)^3.
\label{eq:Effective}
\eeq
where $\rho_1$ is the monomer material density.
Equations~(\ref{eq:DynamicShapeFactor}, \ref{eq:Effective}) lead to
$\rho_{\textrm{eff}} = \rho_1 / \chi_N^{3}$.

\subsection{Dependence of the hydrodynamic radius on the mean number of nearest neighbours}

The cluster hydrodynamic radius may be expressed in terms of the cluster average monomer
shielding factor $\eta_N$ or the individual $i$th monomer shielding factor
$\eta_{N,i}$ via~\cite{LorenzoJCIS11}
\beq
\frac{R_h}{R_1} = N \, \eta_N = \sum_{i=1}^N \eta_{N,i}.
\label{eq:Defineeta}
\eeq
The orientationally-averaged shielding factor, either average or individual, not only allows
the explicit calculation of the hydrodynamic radius, but it has also been used to calculate
a cluster's permeability and thereby its hydrodynamic radius~\cite{sonntag, Vanni00}.
Short-range within-cluster interactions, which affect
monomer shielding, were incorporated in the calculation of a cluster's
permeability through the individual monomer local coordination number, i.e.,
the number of nearest neighbours of each monomer
(number of  touching monomers). Long-range effects were expressed in terms
of the average volume fraction.

We used collision-rate simulations to calculate the average
shielding factor. Results are shown in Table~\ref{table:eta}.
We note that $\eta_N$ depends not only on short-range effects, as
modelled by the prefactor, but also on long-range effects, as described by the
number of monomers and the fractal dimension. This
observation is further supported by Eq.~(\ref{eq:fitRmRg1}), where
the importance of the number of monomers is explicit.
\begin{table}[htb]
\caption{Average cluster shielding factor of clusters calculated by collision-rate simulations.}
\label{table:eta}
\begin{center}
\begin{tabular}{cccccc} \hline \hline
$N$ & $d_{f}$ & $k_{f}$ & $R_g/R_1$  & $c_N$, Eq.~(\ref{eq:CoordinationNumberVanni}) & $\eta_{N}$ \\ \hline
16 & 1.8 & 1.3  & 4.03 & 1.875 & 0.239 \\
32 & 1.8 & 1.6  & 5.28 & 1.937 & 0.157 \\
32 & 1.8 & 1.3  & 5.93 & 1.937 & 0.171 \\
32 & 1.5 & 1.3  & 8.46 & 1.937 & 0.198 \\ \hline \hline
\end{tabular}
\end{center}
\end{table}
%

\begin{figure}[htb]
\begin{centering}
\includegraphics[width=0.5\columnwidth,height=0.4\columnwidth]{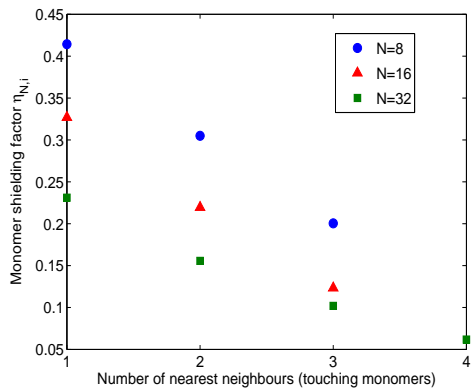}
\caption{Individual-monomer shielding factors within DLCA clusters as a function of
nearest neighbours (local coordination number).}
\label{fig:etaN}
\end{centering}
\end{figure}

The influence of the \textit{local} coordination number, namely
the number of neighbours a chosen monomer has (and not the cluster average),
on the average monomer shielding
within an aggregate was further investigated by calculating the shielding factor of
each monomer in a power-law aggregate.
Figure~\ref{fig:etaN} presents the collision-rate calculated individual-monomer
shielding factors (averaged over very few clusters)
as a function of the number of nearest neighbors
for DLCA clusters
composed of $8,16,32$ monomers. These results, coupled to the
probability distribution of nearest neighbours, may be used to
calculate the average shielding factor.
However, we note that the individual shielding factors do not
fall on a ``universal" (independent of $N$) line, but fall into three lines parametrized
by the number of monomers in the aggregate.
Thus, the $\eta_{N, i}$ shown in Fig.~\ref{fig:etaN}, being dependent on the
overall number of monomers, may not be easily used to estimate the cluster friction coefficient
(or permeability) of clusters composed of an arbitrary number of monomers.
The results are consistent with the previously made observation
that the shielding factor of a monomer depends on a short-range effect,
expressed by the number of nearest neighbours, and a long-range
effect related to the large-scale structure and the size of the cluster,
as noted in Ref.~\cite{Vanni00}.

\section{Conclusions}
\label{sec:Conclusions}

The purpose of our study was to investigate the relationship between structural
and dynamic properties of fractal-like aggregates in the continuum mass
and momentum transfer regimes. We calculated the hydrodynamic radius of synthetic
colloidal aggregates through the numerical solution of a diffusion equation
with appropriate boundary conditions. The resulting normal diffusive flux was related to the
molecule-aggregate collision rate and eventually to the aggregate friction coefficient.
The power-law aggregates used in the simulations were generated via a cluster-cluster
aggregation algorithm.

The morphology of the synthetic aggregates was analyzed
via the three-monomer angle distribution, the mean number of nearest neighbours
(monomers in the first coordination shell), and
the distribution of cluster shape anisotropy. The large-scale distribution of monomers within a
power-law aggregate
is mainly determined by the fractal dimension $\df$, even though for fixed $d_f$,
the mean shape anisotropy
provides a good descriptor of aggregate morphology at large scales.
The fractal prefactor $\kf$, dependent on the shape anisotropy, describes the
local monomer distribution, as determined by the three-monomer angle distribution and
the average number of nearest monomer neighbors.

The aggregate hydrodynamic radius $R_h$, equal to the mobility radius
in the continuum regime, was related to the radius
of gyration $\Rg$ and the number of primary particles (monomers) $N$
via an empirical formula leading to $R_h(N,R_g(\df, \kf, N); R_1)$.
The suggested relationship shows the importance of both $\df$ and $\kf$ in determining the dynamics
of an aggregate; however, their individual values are not required separately
since the hydrodynamic radius may be predicted through their combined
effect as specified by the radius of gyration.
Furthermore, since the proposed
expression does not depend on statistical
cluster properties (like $\df$ and $\kf$) it may be used to estimate the hydrodynamic radius
of single fractal-like objects. Predictions of the suggested expression were in excellent
agreement with literature values for a large range of different ($\df, \kf$) pairs,
and for aggregates generated by different methods, e.g., a ``mimicking" algorithm or
Langevin dynamics. These comparisons
suggest that the validity of the expression is general enough to be used in different settings.

The hydrodynamic-radius expression was used to study the scaling law connecting the number
of monomers to the hydrodynamic radius of DLCA ($1.8, 1.3$) and RLCA ($2.05, 0.94$) clusters.
We found, in agreement with previous works, that the fractal exponent determined
from the radius of gyration and the mass-mobility fractal dimension
determined from the hydrodynamic radius differed, suggesting
that the hydrodynamic radius is not a linear function of the radius of gyration
(as manifested by the proposed expression).

We, also, calculated the shielding factor of individual monomers in
DLCA aggregates. Since the fractal dimension and prefactor were taken to be
constant for DLCA clusters,
the effect of the number of primary particles and dprobability
distribution of nearest neighbours (considered as
an indicator of a cluster's small-scale morphology) were studied. We found
that the individual shielding factor, and consequently the cluster's hydrodynamic behaviour,
depends on the combined effect of small- and large-scale structural properties,
since both the number of nearest neighbours (local structure) and primary particles (large-scale
structure) influence the shielding factors.
Consequently, the pair ($\df , \kf$) is required for a full characterization of both the
structure and dynamics of a power-law aggregate.

\end{document}